\documentclass[cits]{PoS}                                                                                                                                                       
\usepackage{amsmath, amssymb, bm}                                                                                                                                               
\usepackage{subfigure}
\usepackage{multirow}
\usepackage{macros}
\usepackage{wrapfig}

\newcommand{\preprintline}{\newline
\rightline{\parbox{5cm}{\large\tt CERN-PH-TH-2011-314\\SFB/CPP-11-78\\DESY 11-241\\HU-EP-11/60\\}}
}

\title{Strange quark mass and Lambda parameter by the ALPHA collaboration}                                                                                                
                                                                                                                                                                          
\ShortTitle{Strange quark mass and Lambda parameter by the ALPHA collaboration}                                                                                           
                                                                                                                                                                          
\author{\speaker{Marina Marinkovic}\\
	    Humboldt Universit\"at zu Berlin, Institut f\"ur Physik,        
	    Newtonstr. 15, 12489 Berlin, Germany\\                                                                                              
	    E-mail: \email{marina.marinkovic@physik.hu-berlin.de}}                                     
              
\author{Stefan Schaefer\\
	CERN, Physics Department, 1211 Geneva 23, Switzerland \\ 
        E-mail: \email{stefan.schaefer@cern.ch}}

\author{Rainer Sommer, Francesco Virotta\\
	NIC, DESY, Platanenallee 6, 15738 Zeuthen, Germany\\
        E-mail: \email{Rainer.Sommer@desy.de, Francesco.Virotta@desy.de}}

\abstract{We determine $\fK$ for lattice QCD in the two flavor approximation with non-perturbatively 
improved Wilson fermions. The result is used to set the scale for dimensionful quantities in CLS/ALPHA
simulations. To control its dependence on the light quark mass, two different strategies for the chiral
extrapolation are applied. Combining $\fK$ and the bare strange quark mass with non-perturbative 
renormalization factors and step scaling functions computed in the Schr\"odinger Functional, we determine 
the RGI strange quark mass and the Lambda parameter in units of $\fK$.\\
~\\
~\\
\preprintline
}                                                                                                                                    
                                                                                                                                                                          
\FullConference{XXIX International Symposium on Lattice Field Theory \\                                                                                                   
                 July 10-16 2011\\                                                                                                                                      
                 Squaw Valley, Lake Tahoe, California}                                                                                                                    
                                                                                                                                                                          
\begin{document}                                                                                                                                                          
\section{Introduction}                                                                  
\label{s:intro}
The determination of the fundamental parameters of the standard model has a long tradition 
in lattice QCD. In
particular the quark masses and the scale parameter $\Lambda$ can be determined from
first principles. This study is a part of a long-term programme of the ALPHA collaboration 
of computing these parameters, using the
Schr\"odinger functional strategy to overcome the multi-scale problem
and keep the full control over the systematic errors. 

The new ingredient presented here is the scale setting 
using a physical quantity, the Kaon decay constant $\fK$.
With this scale we achieve a 5\% error, employing two different
strategies for the  chiral extrapolation which agree within errorbars.
This enables us to give physical values for the RGI values of the strange quark mass and 
$\Lambda$-parameter, in the setup with two dynamical flavors of light quarks.

The differences to the previously published values for $\Msr$\cite{DellaMorte:2005kg} and 
$\Lambda$\cite{DellaMorte:2004bc}, for $\Nf=2$, come from the
improved scale setting. In the old computation, we used values available
from the literature\cite{Gockeler:2004rp}, where the scale was set with $r_0$. More recent 
determinations of $r_0/a$ find somewhat different results\cite{Leder:Latt2011,Pleiter:Latt2011}. 
Additional improvement comes from lattices with smaller pion masses and
finer lattice spacing than previously available, giving  a better handle
on systematic effects. They were 
generated by the ALPHA Collaboration and the  CLS\footnote{Coordinated
Lattice Simulations} effort.
\section{Action and algorithms}
\label{s:acalg}
Our study is based on ensembles generated with the Wilson plaquette gauge action together 
with $\Nf=2$ mass-degenerate flavors of $O(a)$ improved Wilson fermions. The simulations 
are using either M. L\"uscher's implementation of the DD-HMC algorithm\cite{Luscher:2005rx} 
or our implementation of the MP-HMC algorithm\cite{Marinkovic:2010eg}.

The list of ensembles used in the analysis is shown in Table \ref{table:CLS}. Lattice 
spacings are ranging from 0.05fm to 0.08fm and their precise determination will be presented 
in the following section. The ensembles cover a wide range of pion masses going down to 
$270\mathrm{MeV}$, whereas all lattice volumes satisfy the requirement $\mp L \ge 4$ 
to keep finite volume effects under control.
\begin{table}
\begin{center}
\begin{small}
\begin{tabular}[b]{| c | c | c | c | c | c |c |}
\hline
$~$ & ${\kappa}_{{sea}}$ & ${\mp [\mathrm{MeV}]} $ & ${\mp} L$
& $\mathrm{MDU}$ & $\tint ({\mp})[\mathrm{MDU}]$ & $R_{\mathrm{act}}\texp$ \\[1mm]\hline
\multirow{3}{*} {${\beta}$ = 5.2~~~~}& 0.13565 &  632(20)  & 7.7 & 2950 & 10 &\\[1mm] 
&0.13580 & 495(16) &  6.0  & 2950 & 6& \\[1mm]
&0.13590 & 385(13) &  4.7  & 2986 & 5& 25 \\[1mm]
\multirow{1}{*} {${a} \sim 0.08\mathrm{fm}$}& 0.13594 & 331(11) &  4.0  & 3094 &  5&\\[1mm]\hline
\multirow{3}{*}{${\beta}$ = 5.3~~~~} & 0.13610  & 582(10) & 6.2 & 927 & 18&\\[1mm]
& { 0.13625}  &   437(7)& { 4.7 }&  5900 &  9&\\[1mm]
& 0.13635  & 312(5)& 5.0 & 1769 &  8 & 50\\[1mm]
\multirow{1}{*}{${a}\sim0.07\mathrm{fm}$ }&  0.13638  &  267(5)  &  4.2 & 3473  &  7&\\[1mm]\hline 
\multirow{2}{*}{${\beta}$ = 5.5~~~~} & 0.13650 & 552(6)   & 6.5&1661 & 34 &\\[1mm]
&  0.13660  & 441(5)  & 5.2 & 1686 & 30 & 200\\[1mm]
\multirow{1}{*}{${a}\sim0.05\mathrm{fm}$}  & 0.13671  & 268(3)  & 4.2 & 2796 & 20 &\\ \hline 
\end{tabular}\end{small} \end{center}
\vspace*{-0.3cm}
\caption{$\Nf=2$ ensembles used in the analysis. $\mathrm{MDU}$ is the number of molecular dynamics 
units of the parts of the run chains used in the analysis. The configurations are saved after every 
4 $\mathrm{MDU}$s. As an illustration of autocorrelations, we give the integrated autocorrelation 
time of the pion mass expressed in $\mathrm{MDU}$. $\texp$ is estimated from $\beta=5.3$ 
and quenched scaling\cite{Schaefer:2010hu} and $R_{{act}}$ is the fraction of active 
links\cite{Luscher:2005rx,Marinkovic:2010eg}.}
\vspace*{-0.3cm}
\label{table:CLS}
\end{table}
\section{Scale setting with $\fK$}
To determine the scale and match to experimental values we have to extrapolate decay constants to
the physical quark masses. For this we use two variants based on chiral perturbation theory(ChPT).
The first one employs SU(3) chiral perturbation theory with a quenched 
strange quark. 
The aim here is to minimize the chiral corrections by keeping the sum $\hat{M}+\Ms$
of the light quark mass and the strange quark mass approximately fixed. Chiral corrections are 
expected to be well behaved, since in this setup all Goldstone bosons have a mass of at most the 
physical kaon mass ($500\mathrm{MeV}$). The second approach uses heavy meson chiral perturbation 
theory (HMChPT), expanding only in the light quark mass ($\hat{M}=(\Mu+\Md)/2$). Whereas the first 
strategy is most useful for $\Nf=2$, the second one is equally well applicable in $\Nf=2$ with a 
quenched strange quark and in the $\Nf=2+1$ theory, the only difference being the low energy 
constants. The difference between the two strategies in approaching the physical point is 
illustrated in Figure \ref{Strat1}.

In the setup described in the next two sections we use two 
quarks with hopping parameters 
${\kappa_1} = \kappa_2 = \hat{\kappa}$ 
and two additional quenched quarks with hopping parameters $\kappa_3=\kappa_4$. 
\begin{wrapfigure}{r}{0.45\textwidth}
\begin{center}
\vspace*{-1.2cm}
\includegraphics[scale=0.27,angle=270]{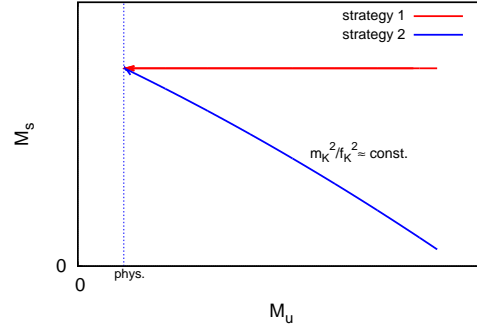}
\label{Strat1}
\vspace*{-0.5cm}
\end{center}
\caption{A sketch of the two approaches for chiral extrapolation to the physical point. 
Strategy 1 imposes the condition on the sum of the strange and light quark 
mass: $\Ms+\hat{M} =const. + O(M^2)$, where $\hat{M}=(\Mu+\Md)/2$, while the second strategy 
keeps the strange quark mass $\Ms$ constant during the extrapolation.}
\label{Strat1}
\vspace*{-0.4cm}
\end{wrapfigure}
From these we build pseudoscalars, pions with mass $\mp$ from two quarks with ${\kappa}_1$ 
and ${\kappa}_2$. The kaons we build from $(\hat{\kappa},\kappa_3)$. The physical point is defined 
by $\mpp=134.8 \mathrm{MeV}$ and $\mKp=494.2 \mathrm{MeV}$, 
the values in QCD with the electromagnetic interaction being switched off\cite{Colangelo:2010et}.
The two strategies differ in how $\kappa_3$ is chosen as a function of $\hat{\kappa}$.

In the following computations we have included the effect of the autocorrelations in the error 
analysis in a very conservative way. Namely, for the estimation of the error we take into account the 
tail of the autocorrelation function\cite{Schaefer:2010hu}. Thus, we are convinced that we 
have statistical errors fully under control. The examples of autocorrelation functions for 
$\fpi$ and $\fK$ are shown in Figure \ref{autocorr}. They are computed following the 
procedures detailed in \cite{Wolff}. 
\subsection{SU(3) Chiral Perturbation Theory (Strategy 1)}
In this approach we define the strange quark hopping parameter $\kappa_3$ through the 
dimensionless ratio
\begin{wrapfigure}{r}{0.55\textwidth}
\begin{center}
\vspace*{-2.0cm}
\hspace*{-0.3cm}
\includegraphics[scale=0.35,angle=270,keepaspectratio=]{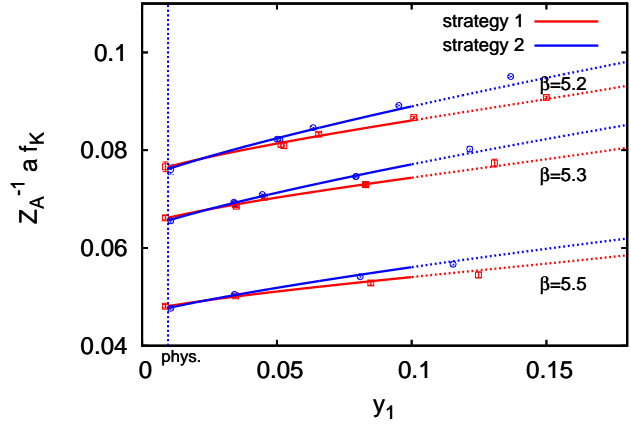}
\label{Strat2}
\vspace*{-0.9cm}
\end{center}
\caption{Chiral extrapolation of the kaon decay constant in lattice units. The values of $\fK$ 
have been multiplied by $Z_A^{-1}$, but the error shown does not take the statistical error 
of the renormalization constant into account. The extrapolated results coming from both 
strategies agree at the physical point. They are obtained by applying the global fit for all 
three values of gauge couplings $\beta$, imposing a cut $\hat{y}_{\pi}<0.1$.}
\label{Strat2}
\vspace*{-0.0cm}
\end{wrapfigure}
\begin{equation}
\frac{\mK^2(\hat{\kappa},\kappa_3)}{\fK^2(\hat{\kappa},\kappa_3)} 
=\frac{\mKp^2}{\fKp^2},
\label{ratio}
\end{equation}
for each value $\hat{\kappa}$ and the gauge coupling ($\beta$), 
such that the l.h.s. of the equation remains equal to the constant 
$R=\frac{\mKp^2}{\fKp^2}$. 
Rather than a fixed strange quark mass, this corresponds to $\Ms+\hat{M}=const.$ to lowest 
order in the expansion in the quark masses and this is expected to give a flat chiral 
extrapolation for $\fK$. The condition (\ref{ratio}) determines a value of 
$\kappa_3=h(\hat{\kappa})$ 
as a function of the sea quark hopping parameter and it can be obtained by interpolation. 
After the dependence of $\kappa_3$ on $\hat{\kappa}$ is determined, it remains to extrapolate 
the decay constant  $a\fK(\hat{\kappa},h(\hat{\kappa}))$ to the physical point, defined by 
the dimensionless ratio
\begin{equation}
\frac{\mp^2(\hat{\kappa},h(\hat{\kappa}))}{\fK^2(\hat{\kappa},h(\hat{\kappa}))} 
=\frac{\mpp^2}{\fKp^2}.
\label{ratioP}
\end{equation}
In the last step we use the prediction of this functional form coming from 
SU(3) ChPT\cite{Sharpe:1997by}:
\begin{eqnarray}
{a}\fK(\hat{\kappa},h(\hat{\kappa}))~&=&~{a}f_{\mathrm{K,lat}}\big[ 1+\bar{L}_\mathrm{K}(\yhp,\yK)+
({\alpha_4}-\frac{1}{4})(\yhp-\yp)+O(y^2) \big],\\
\bar{L}_\mathrm{K}(\yhp,\yK)&=&-\frac{1}{2}\yhp \log (\yhp) - \frac{1}{8} \yhp \log(\frac{2\yK}{\yhp}-1) + 
\frac{1}{2}\yp \log (\yp) 
+ \frac{1}{8} \yp \log (\yp), 
\end{eqnarray}
where ${a}f_\mathrm{K,lat}$ is the value of the decay constant in lattice units and 
the variables $y$ are defined as
\begin{equation}
\yhp=\frac{\mp^2(\hat{\kappa})}{8 \pi^2 \fK^2(\hat{\kappa})} ~~~~~~ 
\yK=\frac{\mKp^2}{8 \pi^2 \fKp^2} ~~~~~~
\yp=\frac{\mpp^2}{8 \pi^2 \fKp^2}.
\end{equation}
The described chiral extrapolation to the physical point is shown in 
Figure \ref{Strat2}. Finally, the lattice spacings for each value of the gauge coupling 
can be obtained with
\begin{equation}
a=\frac{af_{\mathrm{K,lat}}}{\fKp}
\end{equation}
and its values, together with the errors of this determination, are shown in Table 
\ref{latspacStrat1}.
\begin{figure}
\begin{minipage}[b]{0.40\textwidth}
\begin{center}
\includegraphics[scale=0.20,angle=270,keepaspectratio=]{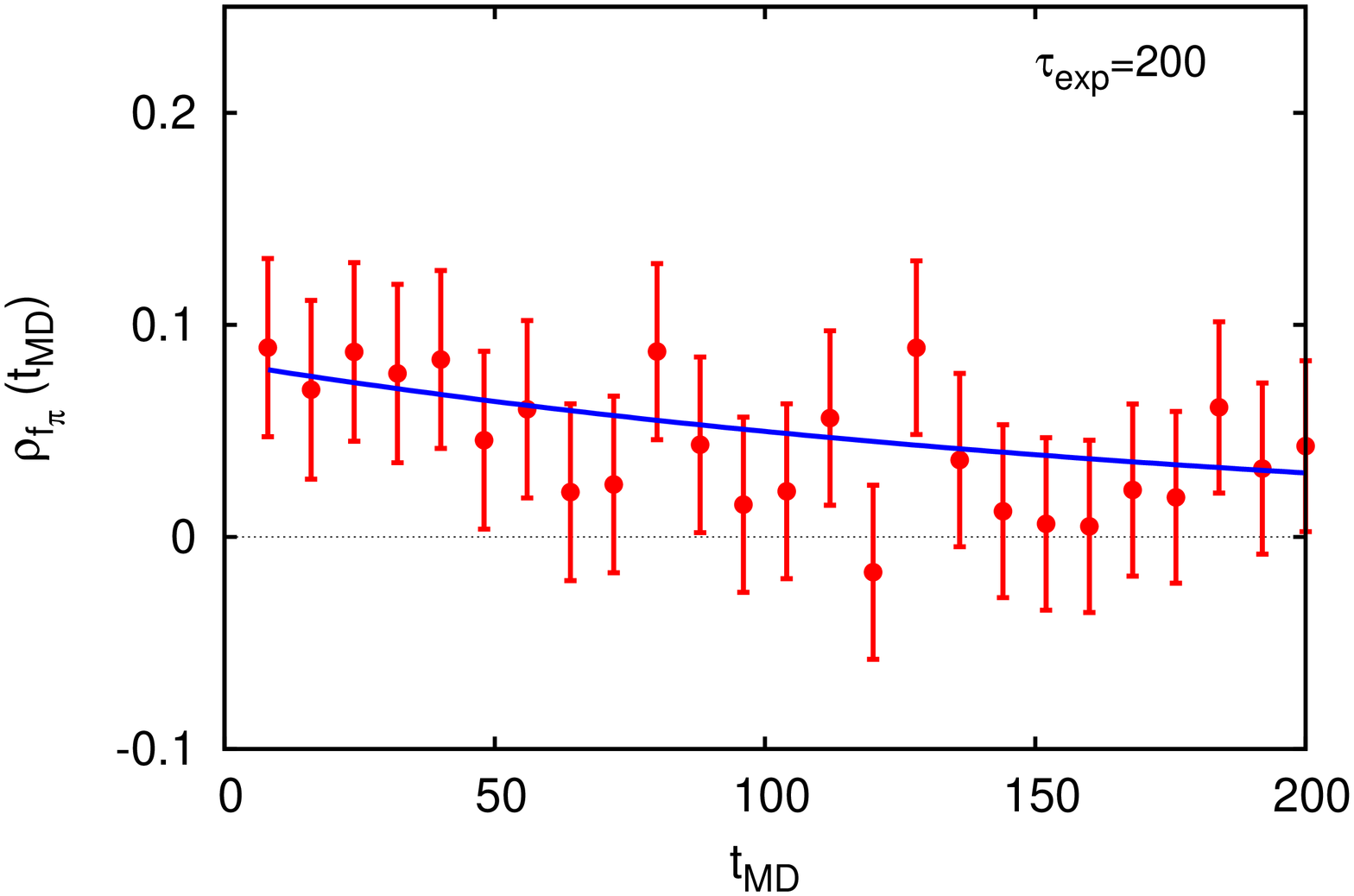}
\end{center}
\end{minipage}\hspace*{0.1\textwidth}
\begin{minipage}[b]{0.40\textwidth}
\begin{center}
\includegraphics[scale=0.20,angle=270,keepaspectratio=]{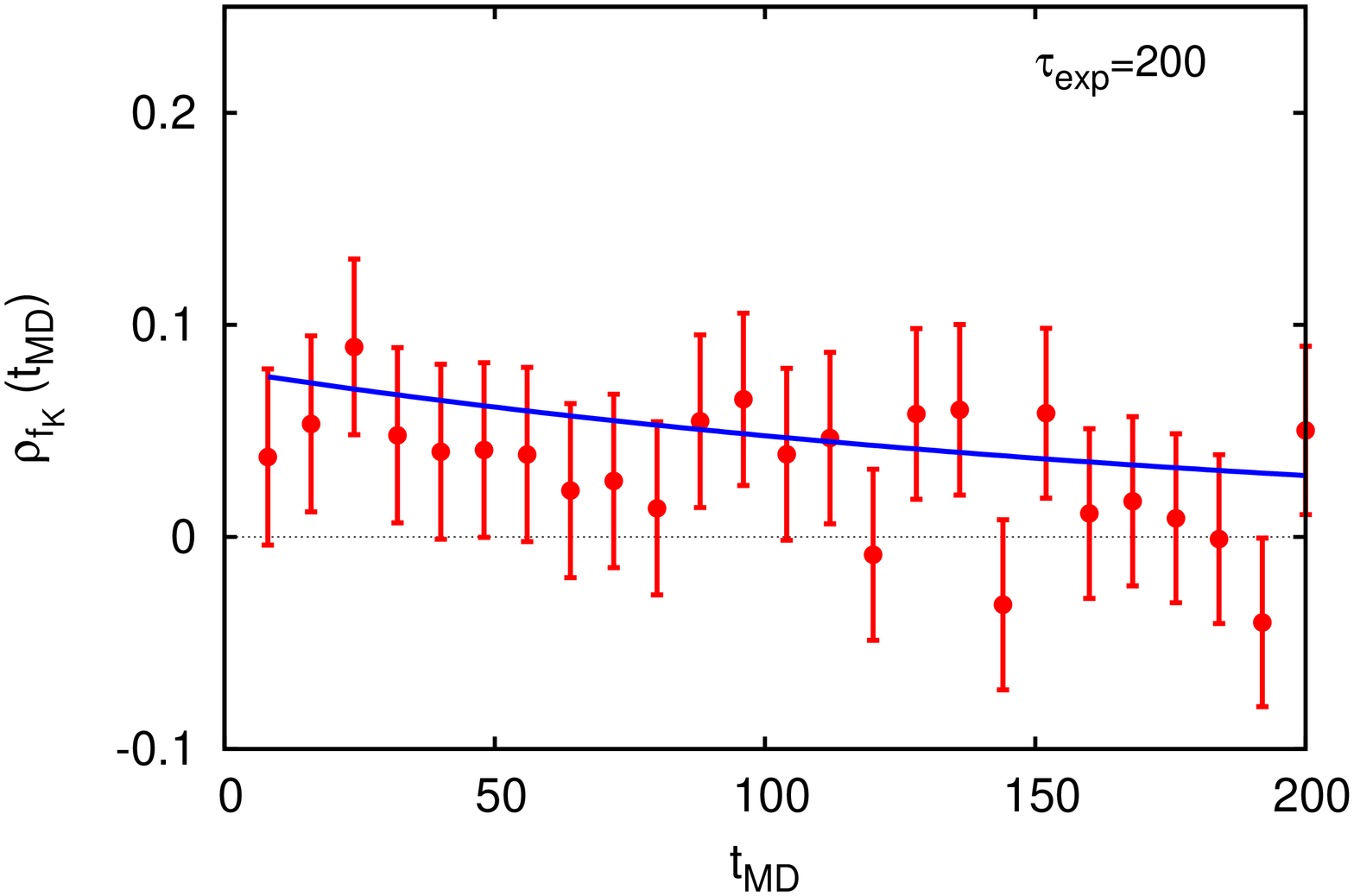}
\end{center}
\end{minipage}
\label{autocorr}
\caption{
Autocorrelation functions of pion decay constant(left) and kaon decay constant(right).
In the error computation the tail of the autocorrelation function is taken into account.
Standard truncations of the integral lead to an error smaller by a factor 2 or more.
}
\label{autocorr}
\end{figure}
\subsection{Heavy Meson Chiral Perturbation Theory (Strategy 2)}
\label{strat2}
In this approach, we work at the fixed strange quark mass and perform a chiral extrapolation 
in the light quark mass ($\hat M$) using HMChPT \cite{Bhattacharya:2005rb}. Since we are 
working with Wilson fermions, one has to keep fixed the axial Ward identity (PCAC) mass of 
the strange quark. If we denote
\begin{equation}
{m}_{{ij}}~=~\frac{1}{2} \frac{\langle\tilde{\partial_{0}}~A^{ij}_{0,\mathrm{I}}~
{{P^{ji}}}\rangle}{\langle{P^{i j}}~{{P^{j i}}}\rangle},
\end{equation}
with ${P^{ij}}=\bar{\psi_{i}}(x) \gamma_5 {\psi_{j}}(x)$ and $A^{ij}_{0,\mathrm{I}}(x)=
\bar{\psi_{i}}(x) \gamma_{0} \gamma_{5} {\psi_{j}}(x) + {a}~\cA\tilde{\partial_{0}}{P^{ij}}(x)$ 
being pseudoscalar density and improved axial current, then the bare PCAC masses of the sea 
and the valence quark are defined by
\begin{equation}
m_1(\hat{\kappa})~=~m_{12}~~~~~~~  m_3(\hat{\kappa},\kappa_3) ~=~m_{34}.
\end {equation}
We first interpolate in $\kappa_3$ and determine functions $s(\hat{\kappa},\mu)$ such that the 
strange quark mass is kept fixed to $m_3(\hat{\kappa},s(\hat{\kappa},\mu))=\mu$. For fixed 
$\mu$ we then perform a HMChPT extrapolation to the chiral limit defined
by $\mpp^2 / \fKp^2$ \cite{Roessl:1999iu,Allton:2008pn}
using
\begin{eqnarray}
{a}~\fK(\hat{\kappa},s(\hat{\kappa},\mu))~&=&~P(\mu) \big[1-\frac{3}{8} [\yhp \log(\yhp)
-\yp \log (\yp)]
+ {\alpha_\mathrm{H}}(\yhp-\yp)+O(M^2)  \big]\\
{a^2}\mK^2(\hat{\kappa},s(\hat{\kappa},\mu))~&=&Q(\mu)\big[1+\alpha'_\mathrm{H}(\yhp-\yp)+O(M^2)\big ].
\end{eqnarray}
In the end, the scale is obtained by interpolation in $\mu$ to the
physical strange quark mass
\begin{equation}
a=\frac{P(\muK)}{\fK}~~~\mathrm{at}~~~\frac{Q(\muK)}{P(\muK)^2}=\frac{\mKp^2}{\fKp^2}.
\end{equation}
The values of the lattice spacings from this strategy are shown in Table \ref{latspacStrat1}. 
Comparing to the results of the first strategy, we find a very good agreement as demonstrated 
in Figure~\ref{Strat2}.
\begin{table}[hd]
\begin{center}
\begin{small}
\begin{tabular}[h]{| c |c|c | c| c| c | c | c|c |}
\cline{3-5}\cline{7-9}
\multicolumn{2}{c|}{~}&\multicolumn{3}{|c|}{Strategy 1}&\multicolumn{1}{|c|}{~}&\multicolumn{3}{|c|}{Strategy 2}\\
\cline{1-1}\cline{3-5}\cline{7-9}
$\beta$ &~& 5.2 & 5.3  & 5.5 &~& 5.2 & 5.3  & 5.5 \\
\cline{1-1}\cline{3-5}\cline{7-9}
$a[\mathrm{fm}]$ &~& 0.0750 &  0.0655  & 0.04847 &~&  0.0745  & 0.0649 & 0.04808 \\ 
\cline{1-1}\cline{3-5}\cline{7-9}
$\Delta_\mathrm{stat.} a$ &~& 0.0024  & 0.0010  & 0.00048 &~&  0.0025 & 0.0010 & 0.00047  \\ 
\cline{1-1}\cline{3-5}\cline{7-9}
$\Delta_\mathrm{syst.} a$ &~&  0.0013  &  0.0011 &  0.00079&~&  0.0014 &  0.0012 & 0.00090 \\
\cline{1-1}\cline{3-5}\cline{7-9}
\end{tabular}
\end{small} 
\end{center}
\vspace*{-0.3cm}
\caption{Lattice spacings from the first strategy obtained by applying SU(3) ChPT(left) 
and from the second strategy, based on HMChPT(right). Estimation of the systematical 
errors is preliminary.}
\label{latspacStrat1}
\vspace*{-0.2cm}
\end{table}
\section{Determination of $\Lambda$ and $\mb_\mathrm{s}$}
In the $\Nf=2$ theory,
low and high energy physics have been connected
non-perturbatively by the ALPHA Collaboration, using an intermediate (Schr\"odinger functional) 
renormalization scheme\cite{DellaMorte:2005kg,DellaMorte:2004bc}. Here QCD is formulated 
in a finite box of spatial size $L$ and temporal extent $T$. The fields are subject to Dirichlet 
boundary conditions in time and periodic in space, where the former provide an infrared cutoff 
to the modes of quarks and gluons. This allows to perform simulations at zero quark mass 
and thus use the SF as a mass-independent renormalization scheme. We additionally specify that 
$T\equiv L$ and then the renormalization conditions are naturally imposed at the scale $\mu=1/L$.

\begin{figure}
\begin{minipage}[b]{0.33\textwidth}
\includegraphics[width=0.9\textwidth,angle=270]{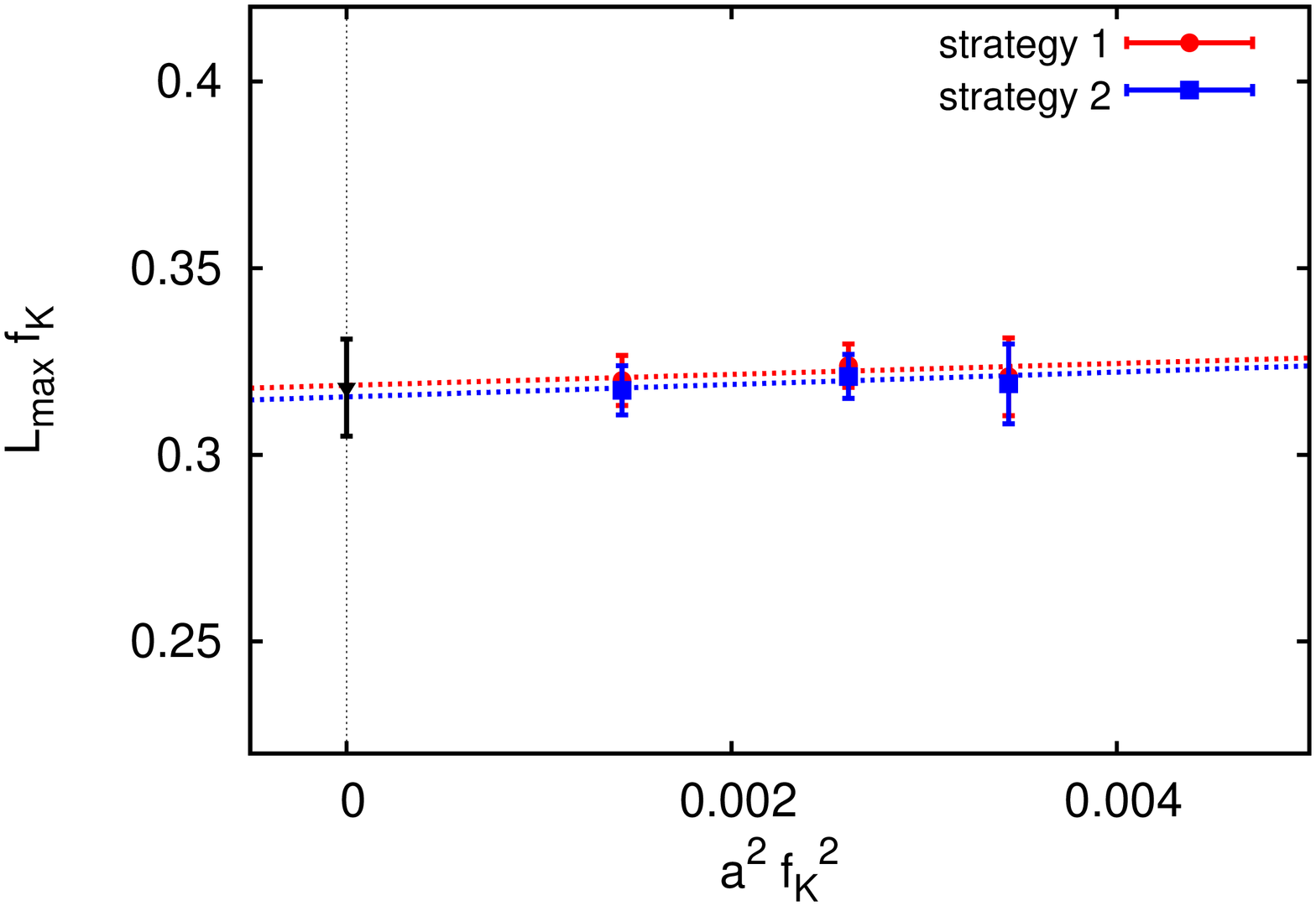}
\end{minipage}
\hspace*{0.15\textwidth}
\begin{minipage}[b]{0.33\textwidth}
\includegraphics[width=0.9\textwidth,angle=270]{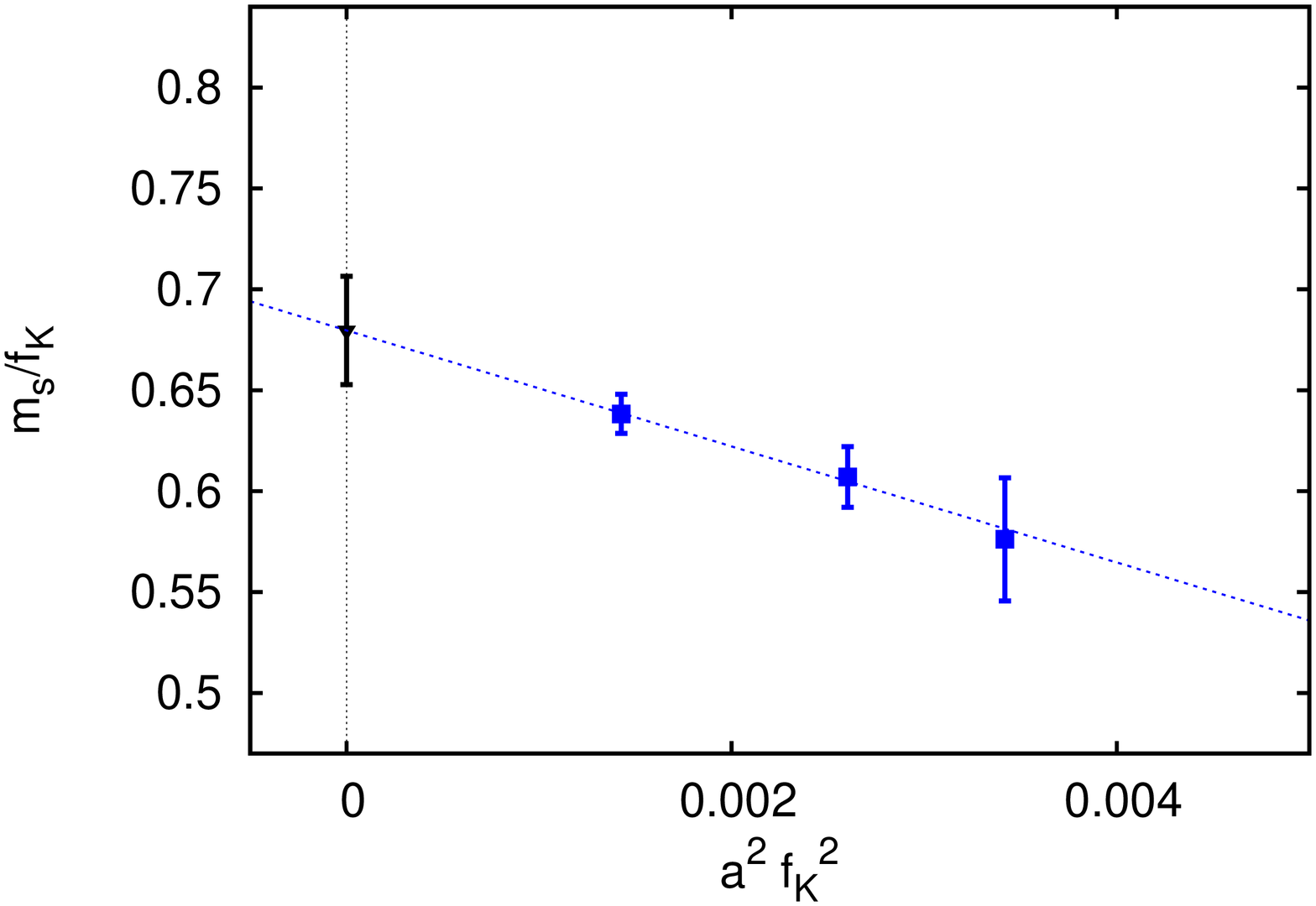}
\end{minipage}
\vspace*{-0.3cm}
\caption{Chiral extrapolations of the product of lattice kaon decay constant 
$\fK$ and matching scale $\Lm$, defined with $\bar{g}^2(\Lm)=4.484$ 
(left) and bare strange quark mass in units of $\fK$(right). The matching scale 
in the extrapolation of $m_\mathrm{s}$(right) is defined with $\bar{g}^2(\Lmt)=4.61$}
\label{lambdams}
\end{figure}

To calibrate the overall energy scale, one fixes a large enough value of the coupling 
$\bar{g}^2(\Lm)$ to be in the low-energy region and relates the associated distance,
 $\Lm$, to a non-perturbative, infinite-volume observable, in our case $\fK$. The extrapolation
of the combination  $\fK \Lm$ 
to the continuum limit is shown in Figure \ref{lambdams}(left). It has been performed for 
both strategies of scale determination and the results agree within the errorbars. Combining 
the continuum result $(\fK \Lm)_\mathrm{cont}~=~0.318(14)(6)$ 
and the value of 
$(\LMSb \Lm)$ 
from \cite{DellaMorte:2004bc}
we get the updated value of the $\MSb$ $\Lambda$-parameter in two flavor QCD:
\begin{equation}
{\Lambda^{(2)}_{\MSb}}={\frac{1}{(\fK \Lm)}} {(\LMSb \Lm)}  \fK  =316(26)(17) \mathrm{MeV},
\end{equation}
where the matching has been performed at the low energy scale $1/\Lm$, defined with 
$\bar{g}^2(\Lm)=4.484$ and $\fK$ is the experimental value of the kaon decay constant.

Furthermore, we compute the strange quark mass. We base it on the PCAC mass
$\mu_\mathrm{K}$ from the second 
strategy for chiral extrapolation (cf. Sect.\ref{strat2}.).
The  $\MSb$ strange quark mass is given by (small corrections
proportional to the quark masses in lattice units are also accounted for)
\begin{equation}
\mb_{\mathrm{s}~\MSb}^{(2)}~(2 \mathrm{GeV})~=
\frac{M}{\mb_\mathrm{SF}(\Lmt)} \, \frac{\za\muK}{\zp(\Lmt)\fK}\, \frac{\mb_{\MSb}(2 \mathrm{GeV})}{M}\,\fK = 
101.4(4.2)(2.5) \mathrm{MeV}, 
\end{equation}
where the first factor is taken over from \cite{DellaMorte:2005kg}, and the new continuum 
extrapolation of the second factor is shown in the right panel of the Figure \ref{lambdams}.
Here, the matching scale is defined with $\bar{g}^2(\Lmt)=4.61$.
The conversion factor to $\MSb$ scheme $\frac{\mb_{\MSb}}{M} (\mu=2\mathrm{GeV})=0.7431$ 
is computed at 4-loop perturbation theory; all other factors are non-perturbative.
\section{Summary and outlook}
An important step missing in our previous work on non-perturbative renormalization of
two flavor QCD has been performed. The scale is set from a physical quantity $\fK$. Two 
strategies of chiral extrapolation are used and both give comparable results. Finally, we 
presented the non-perturbative computation of the $\Lambda$ parameter and the strange 
quark mass of two flavor QCD. The only perturbative input in the whole calculation is 
the 4-loop  
conversion factor from the RGI strange quark mass to $\MSb$ strange quark mass,
which is required to make contact with wide spread conventions.

\section{Acknowledgements}
We thank O.B\"ar, P. Fritzsch, B. Leder, F. Knechtli, H. Simma and U.Wolff for useful and 
stimulating discussions. This work is supported by the German Science Foundation (DFG) 
under the grants GRK1504 "Mass, spectrum, symmetry" and SFB/TR9-03. Simulations were 
performed at JUGENE and JUROPA supercomputers at Forschung Zentrum J\"ulich, HLRN in 
Berlin and Hannover and the PAX Clusters at DESY, Zeuthen.

\end{document}